\documentstyle[12pt,axodraw]{article}
\newcommand{\JHEP}{J. High Energy Phys. }
\newcommand{\RMP}{Rev. Mod. Phys. }
\newcommand{\NP}{Nucl. Phys. }

\newcommand{\PRL}{Phys. Rev. Lett. }
\newcommand{\PL}{Phys. Lett. }

\begin{document}
\baselineskip=20pt

\pagenumbering{arabic}

\vspace{1.0cm}
\begin{flushright}
LU-ITP 2002/005
\end{flushright}

\begin{center}
{\Large\sf Validity of Goldstone theorem at two loops in noncommutative 
$U(N)$ linear sigma model}\\[10pt]
\vspace{.5 cm}

{Yi Liao}
\vspace{1.0ex}

{\small Institut f\"ur Theoretische Physik, Universit\"at Leipzig,
\\
Augustusplatz 10/11, D-04109 Leipzig, Germany\\}

\vspace{2.0ex}

{\bf Abstract}
\end{center}

The scalar theory is ultraviolet (UV) quadratically divergent on 
ordinary spacetime. On noncommutative (NC) spacetime, this 
divergence will generally induce pole-like infrared (IR) 
singularities in external momenta through the UV/IR mixing. 
In spontaneous symmetry breaking theory this would invalidate 
the Goldstone theorem which is the basis for mass generation 
when symmetry is gauged. We examine this issue at two loop level 
in the $U(N)$ linear $\sigma$ model which is known to be free of 
such IR singularities in the Goldstone self-energies at one loop. 
We analyze the structures in the NC parameter 
($\theta_{\mu\nu}$) dependence in two loop integrands of 
Goldstone self-energies. We find that their coefficients are 
effectively once subtracted at the external momentum $p=0$ due to 
symmetry relations between 1PI and tadpole contributions, leaving 
a final result proportional to a quadratic form in $p$. We then 
compute the leading IR terms induced by NC to be of order 
$p^2\ln(\theta_{\mu\nu})^2$ and $p^2\ln\tilde{p}^2$ 
($\tilde{p}_{\mu}=\theta_{\mu\nu}p^{\nu}$) which are much milder 
than naively expected without considering the above cancellation. 
The Goldstone bosons thus keep massless and the theorem holds true 
at this level. However, the limit of $\theta\to 0$ cannot 
be smooth any longer as it is in the one loop Goldstone 
self-energies, and this nonsmooth behaviour is not necessarily 
associated with the IR limit of the external momentum as we see 
in the term of $p^2\ln(\theta_{\mu\nu})^2$. 

\begin{flushleft}


\end{flushleft}

\newpage
\section{Introduction}

Quantum field theory on noncommutative (NC) spacetime may be formulated 
in terms of the Moyal-Weyl correspondence $\cite{review}$. 
Namely, one still works on commutative spacetime but replaces the usual 
product of functions by the star product, 
\begin{equation}
\begin{array}{rcl}
(f_1\star f_2)(x)&=&\displaystyle 
\left[\exp\left(\frac{i}{2}
\theta^{\mu\nu}\partial^x_{\mu}\partial^y_{\nu}\right)
f_1(x)f_2(y)\right]_{y=x}, 
\end{array}
\end{equation}
where $x,y$ are the usual commutative coordinates and 
$\theta_{\mu\nu}$ 
is a real, antisymmetric, constant matrix characterizing the 
noncommutativity of spacetime, 
$[\hat{x}_{\mu},\hat{x}_{\nu}]=i\theta_{\mu\nu}$. 
At the classical action level, while the star product in the 
bilinear terms may be identified with the usual one for 
rapidly decaying functions at spacetime infinity, it does modify 
the interaction terms by introducing a phase which in momentum 
space depends on 
$\theta_{\mu\nu}$ 
and the momenta of fields involved. At the quantum level the phase 
results in a new feature never seen in ordinary field theory, 
the ultraviolet-infrared (UV/IR) mixing
$\cite{mixing}$. The basic mechanism for this occurence may be 
understood as follows. When an otherwise UV divergent 
loop integral is multiplied by a phase depending on both the loop 
momentum $k$ and the external momentum $p$, e.g., 
$\exp(i/2\theta_{\mu\nu}k^{\mu}p^{\nu})$, 
it may become UV convergent due to the rapid oscillation of the 
phase in the UV regime. However, this improvement of the UV 
convergence is effective only for a nonvanishing external momentum
(or more precisely for a nonvanishing NC momentum 
$\tilde{p}_{\mu}=\theta_{\mu\nu}p^{\nu}$). 
The hidden singularity from the UV loop momentum will reappear as 
a new form when the external momentum goes to the IR limit. 
Depending on the degree of divergence of the loop integral, this 
NC IR singularity may be pole-like, logarithmic, etc.

The above NC IR singularity, especially the pole-like one, may 
cause serious problems in NC field theory. It leads to a drastic 
modification to dispersion relation at low energy in 
perturbation theory which may make the theory not well-defined in 
the IR. When going beyond one loop level it may destroy or at 
least make unclear the renormalizability of the theory. Indeed, 
most of explicit model analyses made so far are restricted to the 
one loop level and their renormalization is considered for 
nonexceptional NC external momenta $\cite{u1}-\cite{allorders}$. 
For exceptional ones we would 
have to choose a different subtraction scheme. The real $\phi^4$ 
theory has been examined at two loops $\cite{phi4}$, 
but again the main concern 
is with the UV regime of loop momenta for nonexceptional NC 
momenta. For complex $\phi^4$ theory with spontaneous symmetry 
breaking the IR behaviour becomes important as it is related to 
the issue of whether the Goldstone theorem still holds true on NC 
spacetime, namely, whether the masslessness of Goldstone bosons 
is stable against radiative corrections. This is a starting point 
to all attempts of realistic model building including weak 
interactions $\cite{nogo}$. The complex $U(N)$ $\sigma$ model has 
been studied in this context at one loop in Refs. 
$\cite{campbell}\cite{ruiz}$, 
and it was found that there are no NC IR singularities at all in 
the self-energy of Goldstone bosons so that their masslessness is 
guaranteed at this order: both pole-like and logarithmic ones are 
cancelled in the mass correction due to the delicate relations 
governed by the spontaneously broken symmetries as occuring in the 
commutative theory. This is a surprising result since the scalar 
theory is UV quadratically divergent. It would be highly desirable 
to investigate whether this is a special feature at one loop, or 
more importantly whether NC IR singularities at higher orders, if 
any, endanger the masslessness of Goldstone bosons making the 
theorem no longer valid on NC spacetime. Naively speaking, this 
should not be surprising if it occurs. Beyond one loop, the 
would-be NC IR singularities for external momenta at one loop now 
appear as an internal part of higher loops; it is not clear 
whether they persist to be cancelled. Even worsely, they may 
combine with remaining massless Goldstone bosons to enhance the 
IR behaviour in external momenta. It is the purpose of the present 
work to clarify these problems by an explicit two loop analysis. 
Our main results may be summarized as follows. At two loop level, 
there are no IR terms more singular than $p^2\ln\tilde{p}^2$, 
and individual stronger singularities at intermediate steps are 
finally cancelled due to symmetry relations. The Goldstone bosons 
thus keep massless and the Goldstone theorem holds valid at this 
order in perturbation theory. We also point out the difference 
between the NC singularity in the limit of $\theta_{\mu\nu}\to 0$ 
and the NC IR singularity in the limit of $p\to 0$. For the 
self-energies of Goldstone bosons obtained we have the NC 
behaviour of 
$p^2\ln\theta_{\mu\nu}^2$ and $p^2\ln\tilde{p}^2$. 
While the latter is leading in the IR limit, both are singular in 
$\theta$: we have NC singularities at higher orders independently 
of external momentum configurations.

In the next section we describe the NC $U(N)$ linear $\sigma$ model, 
whose Feynman rules are reproduced in appendix A. Then we present a 
detailed two loop analysis in section 3. Some examples of two loop 
integrals involving $\theta_{\mu\nu}$ are shown in appendix B. 
We conclude in the last section.

\section{The model}

We follow the same conventions as in Ref. $\cite{campbell}$ in 
describing the NC $U(N)$ linear $\sigma$ model. The complex scalar 
$\Phi$ is in the fundamental representation of $U(N)$ with 
\begin{equation}
\begin{array}{rcl}
{\cal L}&=&\displaystyle
(\partial_{\mu}\Phi)^{\dagger}\star\partial^{\mu}\Phi
+\mu^2\Phi^{\dagger}\star\Phi
-\lambda\Phi^{\dagger}\star\Phi\star\Phi^{\dagger}\star\Phi.
\end{array}
\end{equation}
The spontaneous symmetry breaking is triggered by the non-vanishing 
scalar VEV, assuming $\mu^2,\lambda>0$, 
\begin{equation}
\begin{array}{rcl}
\Phi&=&\phi+\phi_0,\\
\phi&=&
\left(\pi_1,\dots,\pi_{N-1},(\sigma+i\pi_0)/\sqrt{2}\right)^T,\\
\phi_0&=&(0,\dots,0,v/\sqrt{2})^T,
\end{array}
\end{equation}
with $v=\sqrt{\mu^2/\lambda}$. The $\sigma$ field is the 
Higgs boson with mass $m=\sqrt{2\lambda v^2}$ and the 
$\pi_0$ and $\pi_i(i=1,\dots,N-1)$ fields are the real and complex 
Goldstone bosons. We have ignored other possible orderings of 
interaction like 
$\Phi^{\dagger}_i\star\Phi^{\dagger}_j\star\Phi_i\star\Phi_j$ 
which are problematic already at one loop 
$\cite{campbell}\cite{ruiz}$.
In terms of the shifted fields, we have 
\begin{equation}
\begin{array}{rcl}
{\cal L}&=&\displaystyle
 \frac{1}{2}(\partial_{\mu}\sigma)^2
-\frac{1}{2}m^2\sigma^2
+\frac{1}{2}(\partial_{\mu}\pi_0)^2
+\frac{1}{2}\partial_{\mu}\pi_i^{\dagger}\partial^{\mu}\pi_i\\
&&\displaystyle 
-\lambda v\sigma (\sigma^2+\pi_0^2+2\pi_i^{\dagger}\pi_i)\\
&&\displaystyle 
-\lambda\left(\frac{1}{4}(\sigma^4+\pi_0^4)
+\pi_i^{\dagger}\pi_i\pi_j^{\dagger}\pi_j\right)\\
&&\displaystyle 
-\lambda\left(\sigma^2\pi_0^2 -\frac{1}{2}\sigma\pi_0\sigma\pi_0
+(\sigma^2+\pi_0^2)\pi_i^{\dagger}\pi_i\right)\\
&&\displaystyle 
-\lambda i[\sigma,\pi_0]\pi_j^{\dagger}\pi_j,
\end{array}
\end{equation}
where we have suppressed the star notation and dropped terms 
which vanish upon integration over spacetime. 
The perturbation theory is based on the above Lagrangian. The one 
loop calculation has been done in Refs. 
$\cite{campbell}\cite{ruiz}$. 
We now proceed to consider two loop contributions in the next 
section. 

\section{Two loop contributions}

There are three sets of contributions at two loop level: bare two 
loop diagrams, one loop diagrams with one insertion of counterterms 
determined at one loop, and the counterterms determined at two loops. 
It is clear that the third causes no IR problem. We start with the 
second which is just a one loop calculation. 

\subsection{One loop diagrams with one insertion of one loop 
counter-terms}

The contributing diagrams are shown in Figs. 1 and 2 where the solid 
and dashed lines are for $\sigma$ and $\pi_{0,i}$ fields 
respectively. As we are 
explicitly including the tadpole contributions, we shall not impose 
the requirement of tadpole cancellation, nor introduce a counterterm 
for the VEV. The counterterms for the self-energies are respectively, 
\begin{equation}
\begin{array}{rl}
\pi_0,\pi_j:&i[p^2\delta Z_{\phi}-m^2\delta_{\pi}]\\
\sigma:&i[(p^2-m^2)\delta Z_{\phi}-m^2\delta_{\sigma}].
\end{array}
\end{equation}
The vertex counterterms are obtained simply by attaching a factor of 
$\delta Z_{\lambda}$ to their Feynman rules. The quantities 
$\delta Z_{\phi,\lambda},\delta_{\pi,\sigma}$ are renormalization 
constants whose details may be found, e.g., in Refs. $\cite{liao}$. 
For our purpose here, it is sufficient to know that
\begin{equation}
\begin{array}{rcl}
\delta_{\sigma}-\delta_{\pi}&=&\delta Z_{\lambda}-\delta Z_{\phi},
\label{eq_01}
\end{array}
\end{equation}
which arises due to their different mass and renormalization.

\begin{center}
\begin{picture}(360,90)(0,0)
\SetOffset(0,0)
\DashCArc(40,45)(15,0,360){3}
\DashLine(10,30)(70,30){3}
\Text(45,0)[]{$(a)$}\Text(40,60)[]{$\times$}
\SetOffset(90,0)
\CArc(40,45)(15,0,360)
\DashLine(10,30)(70,30){3}
\Text(45,0)[]{$(b)$}\Text(40,60)[]{$\times$}
\SetOffset(180,0)
\DashCArc(40,45)(15,0,180){3}\CArc(40,45)(15,180,360)
\DashLine(10,45)(25,45){3}\DashLine(55,45)(70,45){3}
\Text(45,0)[]{$(c)$}\Text(40,60)[]{$\times$}
\SetOffset(270,0)
\DashCArc(40,45)(15,0,180){3}\CArc(40,45)(15,180,360)
\DashLine(10,45)(25,45){3}\DashLine(55,45)(70,45){3}
\Text(45,0)[]{$(d)$}\Text(40,30)[]{$\times$}
\end{picture}
\end{center}
\begin{center}
\begin{picture}(360,90)(0,0)
\SetOffset(0,0)
\DashCArc(40,45)(15,0,360){3}
\DashLine(10,30)(70,30){3}
\Text(45,0)[]{$(e)$}\Text(40,30)[]{$\times$}
\SetOffset(90,0)
\CArc(40,45)(15,0,360)
\DashLine(10,30)(70,30){3}
\Text(45,0)[]{$(f)$}\Text(40,30)[]{$\times$}
\SetOffset(180,0)
\DashCArc(40,45)(15,0,180){3}\CArc(40,45)(15,180,360)
\DashLine(10,45)(25,45){3}\DashLine(55,45)(70,45){3}
\Text(45,0)[]{$(g)$}\Text(25,45)[]{$\times$}
\SetOffset(270,0)
\DashCArc(40,45)(15,0,180){3}\CArc(40,45)(15,180,360)
\DashLine(10,45)(25,45){3}\DashLine(55,45)(70,45){3}
\Text(45,0)[]{$(h)$}\Text(55,45)[]{$\times$}
\end{picture}\\
Fig. 1: 1PI contributions to $\pi_0$ or $\pi_j$ self-energy. 
\end{center}
\begin{center}
\begin{picture}(360,90)(0,0)
\SetOffset(0,0)
\DashCArc(40,55)(10,0,360){3}
\DashLine(10,30)(70,30){3}\Line(40,30)(40,45)
\Text(45,0)[]{$(a)$}\Text(40,65)[]{$\times$}
\SetOffset(90,0)
\CArc(40,55)(10,0,360)
\DashLine(10,30)(70,30){3}\Line(40,30)(40,45)
\Text(45,0)[]{$(b)$}\Text(40,65)[]{$\times$}
\SetOffset(180,0)
\DashCArc(40,55)(10,0,360){3}
\DashLine(10,30)(70,30){3}\Line(40,30)(40,45)
\Text(45,0)[]{$(c)$}\Text(40,45)[]{$\times$}
\SetOffset(270,0)
\CArc(40,55)(10,0,360)
\DashLine(10,30)(70,30){3}\Line(40,30)(40,45)
\Text(45,0)[]{$(d)$}\Text(40,45)[]{$\times$}
\end{picture}\\
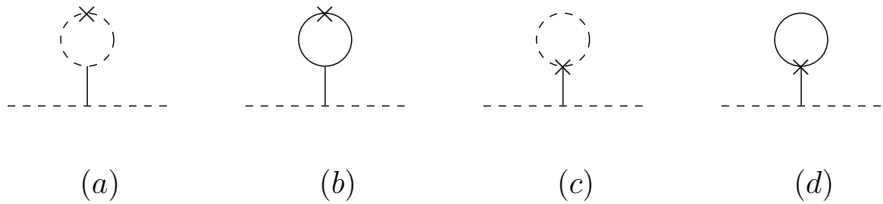
Fig. 2: Tadpole contributions to $\pi_0$ or $\pi_j$ self-energy. 
\end{center}

Let us first consider the part proportional to $\delta Z_{\phi}$.
We have, 
\begin{equation}
\begin{array}{rcl}
[(1a)+(1b)+(1c)+(2a)+(2b)]_{\delta Z_{\phi}}&=&\displaystyle 
({\rm ~one~loop~result~})\times(-\delta Z_{\phi}),
\end{array}
\end{equation}
which is free of NC IR singularities according to Refs. 
$\cite{campbell}\cite{ruiz}$. The remaining  
$\delta Z_{\phi}$ 
dependence will be given below together with that of 
$\delta Z_{\lambda}$. Next, consider the part proportional to 
$\delta Z_{\lambda}$. We have similarly, 
\begin{equation}
\begin{array}{rcl}
[(1e)+(1f)+(1g)+(2c)+(2d)]&=&\displaystyle 
({\rm ~one~loop~result~})\times\delta Z_{\lambda},
\end{array}
\end{equation}
which again is safe. The remaining 
$\delta Z_{\phi}$ and $\delta Z_{\lambda}$ 
dependence is, 
\begin{equation}
\begin{array}{rcl}
[(1d)_{\delta Z_{\phi}}+(1h)]&=&\displaystyle 
\lambda m^2(\delta Z_{\lambda}-\delta Z_{\phi})\\
&&\displaystyle 
\times\left\{\begin{array}{l}
2\delta_{ij}\left[J(0,m)+\cdots\right],
{\rm ~for~}\pi_i^{\dagger}\pi_j\\
\left[J(0,m)+J_{\theta,p}(0,m)+\cdots\right],
{\rm ~for~}\pi_0\pi_0 
             \end{array}
\right.
\label{eq_02}
\end{array}
\end{equation}
where the dots stand for the terms which are both UV (loop momentum) 
and IR (external momentum) finite. Following Ref. 
$\cite{campbell}$ we have introduced similar notations for 
integrals,
\begin{equation}
\begin{array}{rclrcl}
J(0)&=&\displaystyle 
\int\frac{d^4k}{(2\pi)^4}\frac{1}{(k^2)^2},
& 
J_{\theta,p}(0)&=&\displaystyle 
\int\frac{d^4k}{(2\pi)^4}\frac{\cos(2k\wedge p)}{(k^2)^2},\\
J(m)&=&\displaystyle 
\int\frac{d^4k}{(2\pi)^4}\frac{1}{(k^2-m^2)^2},
& 
J_{\theta,p}(m)&=&\displaystyle 
\int\frac{d^4k}{(2\pi)^4}\frac{\cos(2k\wedge p)}{(k^2-m^2)^2},\\
J(0,m)&=&\displaystyle 
\int\frac{d^4k}{(2\pi)^4}\frac{1}{k^2(k^2-m^2)}, 
& 
J_{\theta,p}(0,m)&=&\displaystyle 
\int\frac{d^4k}{(2\pi)^4}\frac{\cos(2k\wedge p)}{k^2(k^2-m^2)}. 
\end{array}
\end{equation}
Our manipulations will be independent of schemes used to regularize 
divergences in the above integrals. Now we compute the 
$\delta_{\sigma,\pi}$ terms and obtain,
\begin{equation}
\begin{array}{rcl}
[(1a)+(2a)]_{\delta_{\pi}}&=&\displaystyle 
+\lambda m^2\delta_{\pi}\left\{
                              \begin{array}{l}
\delta_{ij}2J(0) \\
\left[J(0)+J_{\theta,p}(0)\right] \\
                              \end{array}
\right. \\
\left[(1b)+(2b)\right]_{\delta_{\sigma}}&=&\displaystyle 
-\lambda m^2\delta_{\sigma}\left\{
                                 \begin{array}{l}
\delta_{ij}2J(m) \\
\left[J(m)+J_{\theta,p}(m)\right] \\
                                 \end{array}
\right. \\
(1c)_{\delta_{\pi}}&=&\displaystyle 
+\lambda m^2\delta_{\pi}\left\{
                             \begin{array}{l}
\delta_{ij}2[J(0,m)-J(0)]+\cdots \\
\left[J(0,m)-J(0)+J_{\theta,p}(0,m)-J_{\theta,p}(0)\right]+\cdots \\
                             \end{array}
\right. \\
(1d)_{\delta_{\sigma}}&=&\displaystyle 
-\lambda m^2\delta_{\sigma}\left\{
                                 \begin{array}{l}
\delta_{ij}2[J(0,m)-J(m)]+\cdots \\
\left[J(0,m)-J(m)+J_{\theta,p}(0,m)-J_{\theta,p}(m)\right]+\cdots \\
                                 \end{array}
\right.
\label{eq_03}
\end{array}
\end{equation}
Using eqn. $(\ref{eq_01})$, the IR singularities are cancelled in the 
sum of eqns. $(\ref{eq_02})$ and $(\ref{eq_03})$ leaving behind an 
IR safe result proportional to $p^2$. 

\subsection{Two loop diagrams}

Now we calculate the genuine two loop contributions. The 1PI diagrams 
are depicted in Figs. $3$ and $4$ where we only show topologically 
different graphs with the solid line representing all scalar fields. 
The number appearing as a subscript refers to the number of diagrams 
actually involved.

\begin{center}
\begin{picture}(240,90)(0,0)
\SetOffset(0,0)
\CArc(40,55)(15,0,360)\CArc(40,25)(15,0,360)
\Line(40,90)(40,70)
\Text(40,0)[]{$(a)_9$}
\SetOffset(80,0)
\CArc(40,50)(20,0,360)\Line(20,50)(60,50)
\Line(40,90)(40,70)
\Text(40,0)[]{$(b)_5$}
\SetOffset(160,0)
\CArc(40,50)(20,0,360)
\Line(40,90)(40,30)
\Text(40,0)[]{$(c)_3$}
\end{picture}\\
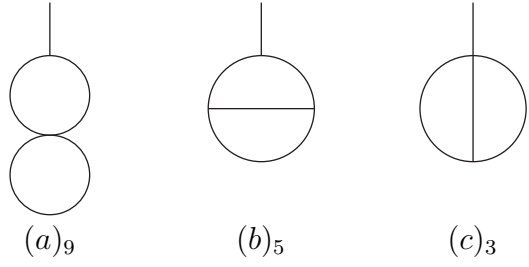
Fig. 3 Two loop 1PI contributions to $\sigma$ tadpole.
\end{center}
\begin{center}
\begin{picture}(360,90)(0,0)
\SetOffset(0,0)
\CArc(40,45)(15,0,360)\CArc(40,75)(15,0,360)
\Line(10,30)(70,30)
\Text(45,0)[]{$(a)_9$}
\SetOffset(90,0)
\CArc(40,50)(20,0,360)
\Line(20,50)(60,50)\Line(10,30)(70,30)
\Text(45,0)[]{$(b)_5$}
\SetOffset(180,0)
\CArc(40,50)(20,0,360)\Line(23,60)(57,60)
\Line(10,50)(20,50)\Line(60,50)(70,50)
\Text(45,0)[]{$(c)_4$}
\SetOffset(270,0)
\CArc(40,50)(20,0,360)\CArc(40,80)(10,0,360)
\Line(10,50)(20,50)\Line(60,50)(70,50)
\Text(45,0)[]{$(d)_6$}
\end{picture}
\end{center}
\begin{center}
\begin{picture}(360,90)(0,0)
\SetOffset(0,0)
\CArc(30,50)(15,0,360)\CArc(60,50)(15,0,360)
\Line(5,50)(15,50)\Line(75,50)(85,50)
\Text(45,0)[]{$(e)_1$}
\SetOffset(90,0)
\CArc(45,50)(20,0,360)
\Line(10,50)(80,50)
\Text(45,0)[]{$(f)_4$}
\SetOffset(180,0)
\CArc(45,50)(20,0,360)\Line(45,70)(45,30)
\Line(10,50)(25,50)\Line(65,50)(80,50)
\Text(45,0)[]{$(g)_2$}
\SetOffset(270,0)
\CArc(25,50)(10,0,360)\CArc(45,40)(25,0,130)
\Line(10,40)(80,40)
\Text(45,0)[]{$(h)_8$}
\end{picture}\\
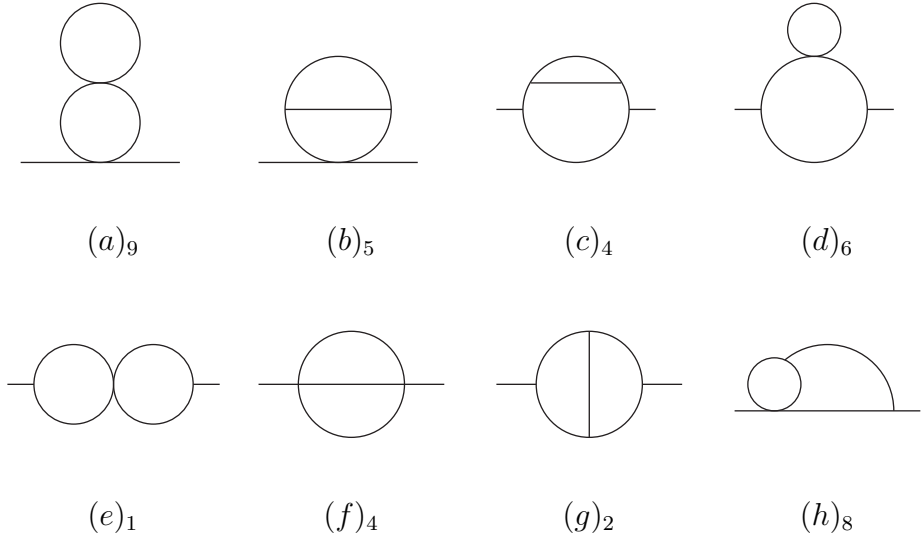
Fig. 4: Two loop 1PI contributions to $\pi_0$ or $\pi_j$ self-energy.
\end{center}

The 1PI $\sigma$ tadpole is found to be, 
\begin{equation}
\begin{array}{rcl}
iT^{\rm 1PI}&=&\displaystyle 
i\lambda^2 v\int\frac{d^4k_1}{(2\pi)^4}\int\frac{d^4k_2}{(2\pi)^4}
T(k_i),\\
T(k_i)&=&\displaystyle 
T_a+T_b+T_c,\\
\end{array}
\end{equation}
where, using the notations 
$D_m(q)=(q^2-m^2)^{-1}$, $D(q)=(q^2)^{-1}$ and 
$K_{12}=\cos(2k_1\wedge k_2)$, we have, 
\begin{equation}
\begin{array}{rcl}
T_a&=&\displaystyle 
+[3D_m^2(k_1)D_m(k_2)+D^2(k_1)D(k_2)](2+K_{12})\\
&&\displaystyle 
+[3D_m^2(k_1)D(k_2)+D^2(k_1)D_m(k_2)](2-K_{12})\\
&&\displaystyle 
+(N-1)[6D_m^2(k_1)D(k_2)+4D^2(k_1)D(k_2)+2D^2(k_1)D_m(k_2)]\\
&&
+4(N-1)ND^2(k_1)D(k_2), 
\end{array}
\end{equation}
\begin{equation}
\begin{array}{rcl}
T_b/m^2&=&\displaystyle 
+[27/2D_m^2(k_1)D_m(k_2)D_m(k_1+k_2)
+3/2D_m^2(k_1)D(k_2)D(k_1+k_2)\\
&&\displaystyle 
+D^2(k_1)D(k_2)D_m(k_1+k_2)](1+K_{12})\\
&&\displaystyle 
+(N-1)[6D_m^2(k_1)D(k_1+k_2)+4D^2(k_1)D_m(k_1+k_2)]D(k_2), 
\end{array}
\end{equation}
\begin{equation}
\begin{array}{rcl}
T_c&=&\displaystyle 
+[3D_m(k_1)D_m(k_2)+D(k_1)D(k_2)]D_m(k_1+k_2)(1+K_{12})\\
&&\displaystyle 
+4(N-1)D(k_1)D(k_2)D_m(k_1+k_2). 
\end{array}
\end{equation}
Note that we can have $\theta$ dependence beyond one loop even if 
the external momentum vanishes since there are independent loop 
momenta which can combine with the antisymmetric $\theta_{\mu\nu}$.
The result will depend on it through 
$\theta^2=\theta_{\mu\nu}\theta^{\mu\nu}$, etc. 
As long as we do not use the Lorentz covariance to choose a 
specific frame for external momenta, we can always treat 
$\theta$ in integrals as if it were a Lorentz tensor. 

Upon choosing loop momenta properly in some integrals, the 1PI 
self-energy of the charged Goldstone bosons $\pi_i^+\pi_j$ is found 
to be, 
\begin{equation}
\begin{array}{rcl}
i\Sigma_{ij}^{\rm 1PI}(p)&=&\displaystyle 
i\lambda^2\delta_{ij}\int\frac{d^4k_1}{(2\pi)^4}\int\frac{d^4k_2}{(2\pi)^4}
U(k_i,p),\\
U(k_i,p)&=&\displaystyle 
\sum_{x=a}^{h}U_x, 
\end{array}
\end{equation}
\begin{equation}
\begin{array}{rcl}
U_a&=&\displaystyle 
+[D_m^2(k_1)D_m(k_2)+D^2(k_1)D(k_2)](2+K_{12})\\
&&\displaystyle 
+[D_m^2(k_1)D(k_2)+D^2(k_1)D_m(k_2)](2-K_{12})\\
&&\displaystyle 
+2(N-1)[D_m^2(k_1)+D^2(k_1)]D(k_2)\\
&&\displaystyle 
+2N[D^2(k_1)D_m(k_2)+D^2(k_1)D(k_2)]\\
&&\displaystyle 
+4N^2D^2(k_1)D(k_2), 
\end{array}
\end{equation}
\begin{equation}
\begin{array}{rcl}
U_b/m^2&=&\displaystyle 
+[9/2D_m^2(k_1)D_m(k_2)D_m(k_1+k_2)+1/2D_m^2(k_1)D(k_2)D(k_1+k_2)\\
&&\displaystyle 
+D^2(k_1)D_m(k_2)D(k_1+k_2)](1+K_{12})\\
&&\displaystyle 
+2(N-1)D_m^2(k_1)D(k_2)D(k_1+k_2)\\
&&\displaystyle 
+4ND^2(k_1)D_m(k_2)D(k_1+k_2), 
\end{array}
\end{equation}
\begin{equation}
\begin{array}{rcl}
U_c/m^4&=&\displaystyle 
+D_m^2(k_1)D(k_1+p)[9D_m(k_2)D_m(k_1+k_2)\\
&&\displaystyle 
+D(k_2)D(k_1+k_2)](1+K_{12})\\
&&\displaystyle 
+4D^2(k_1)D_m(k_1+p)D_m(k_2)D(k_1+k_2)\\
&&\displaystyle 
+4(N-1)D_m^2(k_1)D(k_1+p)D(k_2)D(k_1+k_2), 
\end{array}
\end{equation}
\begin{equation}
\begin{array}{rcl}
U_d/m^2&=&\displaystyle 
+2D_m^2(k_1)D(k_1+p)D_m(k_2)(2+K_{12})\\
&&\displaystyle 
+2D_m^2(k_1)D(k_1+p)D(k_2)(2-K_{12})\\
&&\displaystyle 
+2D^2(k_1)D_m(k_1+p)[D_m(k_2)+D(k_2)]\\
&&\displaystyle 
+4(N-1)D_m^2(k_1)D(k_1+p)D(k_2)\\
&&\displaystyle 
+4ND^2(k_1)D_m(k_1+p)D(k_2), 
\end{array}
\end{equation}
\begin{equation}
\begin{array}{rcl}
U_e/m^2&=&+2D_m(k_1)D(k_1+p)D_m(k_2)D(k_2+p)(1+K_{12}), 
\end{array}
\end{equation}
\begin{equation}
\begin{array}{rcl}
U_f&=&\displaystyle 
+[D_m(k_1)D_m(k_2)+D(k_1)D(k_2)]D(k_1+k_2-p)(1+K_{12})\\
&&\displaystyle 
+2D_m(k_1)D(k_2)D(k_1+k_2-p)(1-K_{12})\\
&&\displaystyle 
+4[(N-1)D(k_1)D(k_2)\\
&&\displaystyle 
+D(k_1+p)D(k_2+p)K_{12}]D(k_1+k_2+p), 
\end{array}
\end{equation}
\begin{equation}
\begin{array}{rcl}
U_g/m^4&=&\displaystyle 
+6D_m(k_1)D(k_1+p)D_m(k_2)D(k_2+p)D_m(k_1-k_2)(1+K_{12})\\
&&\displaystyle 
+4D_m(k_1)D(k_1+p)D(k_2+p)D_m(k_2)D(k_1+k_2+p)K_{12}, 
\end{array}
\end{equation}
\begin{equation}
\begin{array}{rcl}
U_h/m^2&=&\displaystyle 
+D_m(k_1)D(k_1+p)[6D_m(k_2)D_m(k_1+k_2)
+2D(k_2)D(k_1+k_2)\\
&&\displaystyle 
+4D_m(k_2)D(k_1+k_2+p)](1+K_{12})\\
&&\displaystyle 
+8D_m(k_1)D(k_1+p)[(N-1)D(k_2)D(k_1+k_2)\\
&&\displaystyle 
+D(k_2-p)D(k_1+k_2-p)K_{12}]. 
\end{array}
\end{equation}

Although the 1PI self-energy of the neutral Goldstone boson 
$\pi_0$ has the same set of diagrams as the charged one, it 
becomes more complicated due to multiplications of 
trigonometric functions involving the loop and external 
momenta and $\theta$. To simplify our analysis of the NC IR 
behaviour, it is useful to cast these products into standard 
forms. For the self-energy at two loops, we have three 
independent momenta in the integrand, $k_1,k_2$ and $p$ so 
that we can form two independent combinations with 
$\theta$, $2k_1\wedge k_2$ and $2k_1\wedge p$. 
($2k_2\wedge p$ is not independent as it can be obtained 
from $2k_1\wedge p$ by $k_1\leftrightarrow k_2$.) 
We find that it is always possible by shifting and 
interchanging loop momenta properly so that the only 
$\theta$ dependence in integrands enters through either the 
above $K_{12}$ or $K_1=\cos(2k_1\wedge p)$. For example, 
the simple-looking Fig. $4(e)$ in this case involves the 
following product, 
\begin{equation}
\begin{array}{l}
\cos(p\wedge k_1)\cos(p\wedge k_2)\\
\times \{2\cos(k_1\wedge k_2)\cos[(k_1+p)\wedge (k_2+p)]
-\cos[(k_1+k_2)\wedge p]\}\\
=1/4\{\cos(2k_1\wedge k_2)
+\cos[2k_1\wedge k_2+2(k_1-k_2)\wedge p]\\
+\cos[2(k_1+p)\wedge k_2]
+\cos[2(k_2+p)\wedge k_1]\\
+\cos[2(k_1-k_2)\wedge p]
-\cos[2(k_1+k_2)\wedge p]\}, 
\end{array}
\end{equation}
where the five non-standard forms may be transformed into 
the standard ones, e.g., 
\begin{equation}
\begin{array}{rcl}
\cos[2k_1\wedge k_2+2(k_1-k_2)\wedge p]&\to& K_{12},
{\rm ~with~}k_i\to -k_i-p,\\ 
\cos[2(k_1+k_2)\wedge p]&\to& K_1, 
{\rm ~with~}k_1\to -k_1-k_2.
\end{array}
\end{equation}
After this manipulation, the expression for the 1PI $\pi_0$ 
self-energy becomes very lengthy though it has a simpler 
structure in $\theta$,
\begin{equation}
\begin{array}{rcl}
i\Sigma_{00}^{\rm 1PI}(p)&=&\displaystyle 
i\lambda^2\int\frac{d^4k_1}{(2\pi)^4}\int\frac{d^4k_2}{(2\pi)^4}
V(k_i,p)\\
V(k_i,p)&=&\displaystyle \sum_{x=a}^{h}V_x, 
\end{array}
\end{equation}
\begin{equation}
\begin{array}{rcl}
V_a&=&\displaystyle 
+D_m^2(k_1)[D_m(k_2)2(2-K_1+K_{12})-D_m(k_2+p)K_{12}]\\
&&\displaystyle 
+D_m^2(k_1)[D(k_2)2(2-K_1-K_{12})+D(k_2+p)K_{12}]\\
&&\displaystyle 
+D^2(k_1)[D_m(k_2)2(2+K_1-K_{12})-D_m(k_2+p)K_{12}]\\
&&\displaystyle 
+D^2(k_1)[D(k_2)2(2+K_1+K_{12})+D(k_2+p)K_{12}]\\
&&\displaystyle 
+2(N-1)\{[D_m^2(k_1)(2-K_1)+D^2(k_1)(2+K_1)]D(k_2)\\
&&\displaystyle 
+D^2(k_1)[D_m(k_2)+D(k_2)]\}\\
&&\displaystyle 
+4(N-1)ND^2(k_1)D(k_2), 
\end{array}
\end{equation}
\begin{equation}
\begin{array}{rcl}
V_b/m^2&=&\displaystyle 
+9/2D_m^2(k_1)[D_m(k_2)D_m(k_1+k_2)(2-K_1+2K_{12})\\
&&\displaystyle 
-D_m(k_2+p)D_m(k_1+k_2+p)K_{12}]\\
&&\displaystyle 
+1/2D_m^2(k_1)[{\rm same~as~above~except~}m\to 0]\\
&&\displaystyle 
+D^2(k_1)[D_m(k_2)D(k_1+k_2)(2+K_1+2K_{12})\\
&&\displaystyle 
+D_m(k_2+p)D(k_1+k_2+p)K_{12}]\\
&&\displaystyle 
+2(N-1)[D_m^2(k_1)D(k_2)(2-K_1)\\
&&\displaystyle 
+2D^2(k_1)D_m(k_2)]D(k_1+k_2), 
\end{array}
\end{equation}
\begin{equation}
\begin{array}{rcl}
V_c/m^4&=&\displaystyle 
+9/4D_m^2(k_1)D(k_1+p)\{D_m(k_2)D_m(k_1+k_2)2(1+K_1+K_{12})\\
&&\displaystyle 
+[D_m(k_2+p)D_m(k_1+k_2+p)+(p\to -p)]K_{12}\}\\
&&\displaystyle 
+1/2D^2(k_1)D_m(k_1+p)\{{\rm same~as~above~except~}m\to 0
{\rm ~in~each~2nd~}D\}\\
&&\displaystyle 
+1/4D_m^2(k_1)D(k_1+p)\{{\rm same~as~above~except~all~}m\to 0\}\\
&&\displaystyle 
+2(N-1)D_m^2(k_1)D(k_1+p)D(k_2)D(k_1+k_2)(1+K_1), 
\end{array}
\end{equation}
\begin{equation}
\begin{array}{rcl}
V_d/m^2&=&\displaystyle 
+D_m^2(k_1)D(k_1+p)\{[D_m(k_2)(2+2K_1+K_{12})+D_m(k_2+p)K_{12}]\\
&&\displaystyle 
+[D(k_2)(2+2K_1-K_{12})-D(k_2+p)K_{12}]\}\\
&&\displaystyle 
+D^2(k_1)D_m(k_1+p)\{[D_m(k_2)(2+2K_1-K_{12})-D_m(k_2+p)K_{12}]\\
&&\displaystyle 
+[D(k_2)(2+2K_1+K_{12})+D(k_2+p)K_{12}]\}\\
&&\displaystyle 
+2(N-1)\{D_m^2(k_1)D(k_1+p)+D^2(k_1)D_m(k_1+p)\}D(k_2)(1+K_1), 
\end{array}
\end{equation}
\begin{equation}
\begin{array}{rcl}
V_e/m^2&=&\displaystyle 
+[D(k_1+p)D_m(k_1)+D(k_1)D_m(k_1+p)]\\
&&\displaystyle 
\times[D(k_2+p)D_m(k_2)+D(k_2)D_m(k_2+p)]K_{12}\\
&&\displaystyle 
+D_m(k_2)[D(k_2+p)-D(k_2-p)]D_m(k_1+k_2)D(k_1+k_2+p)K_1, 
\end{array}
\end{equation}
\begin{equation}
\begin{array}{rcl}
V_f&=&\displaystyle 
+\{D_m(k_1+p)D_m(k_2)D(k_1+k_2)(3-4K_{12})\\
&&\displaystyle 
-2[D_m(k_1)D_m(k_2+p)+D_m(k_1+p)D_m(k_2)]D(k_1+k_2)K_1\\
&&\displaystyle 
+2D(k_1)D_m(k_2)D_m(k_1+k_2-p)K_1\\
&&\displaystyle 
+[2D_m(k_1)D_m(k_2)+2D_m(k_1-p)D_m(k_2+p)\\
&&\displaystyle 
+D_m(k_1+p)D_m(k_2+p)]D(k_1+k_2+p)K_{12}\}\\
&&\displaystyle 
+1/3\{{\rm same~as~above~except~}-\to +{\rm ~in~first~two~lines~}\\
&&\displaystyle 
{\rm ~and~all~}m\to 0\}\\
&&\displaystyle 
+2(N-1)D(k_2)[D(k_1+p)D(k_1+k_2)+D(k_1)D(k_1+k_2-p)K_1]\\
&&\displaystyle 
+2(N-1)D(k_2)[D(k_1+p)D_m(k_1+k_2)-D_m(k_1)D(k_1+k_2-p)K_1], 
\end{array}
\end{equation}
\begin{equation}
\begin{array}{rcl}
V_g/m^4&=&\displaystyle 
+3/2\{D_m(k_1)D(k_1+p)D_m(k_2)D(k_2+p)D_m(k_1-k_2)(1+K_1+K_{12})\\
&&\displaystyle 
+[D_m(k_1+p)D(k_1)D_m(k_2)D(k_2+p)D_m(k_1+k_2+p)\\
&&\displaystyle 
+D_m(k_1)D(k_1+p)D_m(k_2+p)D(k_2)D_m(k_1+k_2+p)\\
&&\displaystyle 
+D_m(k_1-p)D(k_1)D_m(k_2-p)D(k_2)D_m(k_1-k_2)]K_{12}\\
&&\displaystyle 
+[D_m(k_1)D(k_1+p)D_m(k_2)D_m(k_1-k_2)\\
&&\displaystyle 
+D_m(k_1+k_2)D(k_1+k_2+p)D_m(k_2)D_m(k_1)]D(k_2+p)K_1\}\\
&&\displaystyle 
+1/2\{
{\rm same~as~above~but~interchanging~masses~in~}\\
&&\displaystyle 
{\rm 3rd~and~4th~}D{\rm ~and~}m\to 0{\rm ~in~5th~}D\}, 
\end{array}
\end{equation}
\begin{equation}
\begin{array}{rcl}
V_h/m^2&=&\displaystyle 
+3\{D_m(k_1)D(k_1+p)D_m(k_2)D_m(k_1+k_2)2(1+K_1+K_{12})\\
&&\displaystyle 
+D_m(k_1)D(k_1+p)[D_m(k_2+p)D_m(k_1+k_2+p)+(p\to -p)]K_{12}\\
&&\displaystyle 
-[D_m(k_2)D(k_2+p)D_m(k_1)D_m(k_1+k_2)+(k_2\to -k_1-k_2)]K_1\\
&&\displaystyle 
-D_m(k_1-p)D(k_1)[D_m(k_2)D_m(k_1+k_2-p)+(k_2\to k_2+p)]K_{12}\}\\
&&\displaystyle 
+\{{\rm same~as~above~except~}m\to 0{\rm ~in~3rd~and~4th~}D\\
&&\displaystyle 
{\rm ~and~}-\to + {\rm ~in~last~two~lines}\}\\
&&\displaystyle 
+2\{D(k_1)D_m(k_1+p)[D_m(k_2-p)D(k_1+k_2-p)-(p\to -p)]K_{12}\\
&&\displaystyle 
+D(k_1-p)D_m(k_1)[D_m(k_2)D(k_1+k_2-p)+(k_2\to k_2+p)]K_{12}\\
&&\displaystyle 
+[D_m(k_2)D(k_1)D_m(k_1+k_2+p)\\
&&\displaystyle 
+D_m(k_1)D(k_2)D_m(k_2+p)]D(k_1+k_2)K_1\}\\
&&\displaystyle 
+4(N-1)D_m(k_1)D(k_1+p)D(k_2)D(k_1+k_2)(1+K_1). 
\end{array}
\end{equation}

Including the tadpole contributions, we have the self-energies 
for Goldstone bosons, 
\begin{equation}
\begin{array}{rcl}
i\Sigma_{ij}(p)&=&\displaystyle 
i\lambda^2\delta_{ij}\int\frac{d^4k_1}{(2\pi)^4}\int\frac{d^4k_2}{(2\pi)^4}
[U(k_i,p)-T(k_i)],\\
i\Sigma_{00}(p)&=&\displaystyle 
i\lambda^2\int\frac{d^4k_1}{(2\pi)^4}\int\frac{d^4k_2}{(2\pi)^4}
[V(k_i,p)-T(k_i)], 
\end{array}
\end{equation}
which have the following structure in $\theta$, 
\begin{equation}
\begin{array}{rcl}
i\Sigma_{ij}(p)&=&\displaystyle 
i\lambda^2\delta_{ij}\int\frac{d^4k_1}{(2\pi)^4}\int\frac{d^4k_2}{(2\pi)^4}
[f_0(k_i,p)+f_{12}(k_i,p)K_{12}],\\
i\Sigma_{00}(p)&=&\displaystyle 
i\lambda^2\int\frac{d^4k_1}{(2\pi)^4}\int\frac{d^4k_2}{(2\pi)^4}
[g_0(k_i,p)+g_1(k_i,p)K_1+g_{12}(k_i,p)K_{12}], 
\end{array}
\end{equation}
where all $\theta$ dependence resides in $K$ factors. A crucial 
observation from the above explicit expressions is that all form 
factors $f$ and $g$ vanish at $p=0$ so that we are effectively 
subtracting at $p=0$ for each form factor when doing integrations, 
\begin{equation}
\begin{array}{rcl}
i\Sigma_{ij}(p)&=&\displaystyle 
i\lambda^2\delta_{ij}
\int\frac{d^4k_1}{(2\pi)^4}\int\frac{d^4k_2}{(2\pi)^4}
[\overline{f_0(k_i,p)}+\overline{f_{12}(k_i,p)}K_{12}],\\
i\Sigma_{00}(p)&=&\displaystyle 
i\lambda^2\int\frac{d^4k_1}{(2\pi)^4}\int\frac{d^4k_2}{(2\pi)^4}
[\overline{g_0(k_i,p)}+\overline{g_1(k_i,p)}K_1
+\overline{g_{12}(k_i,p)}K_{12}], 
\end{array}
\end{equation}
where 
$\overline{f_0(k_i,p)}=f_0(k_i,p)-f_0(k_i,0)$, 
$\overline{g_0(k_i,p)}=g_0(k_i,p)-g_0(k_i,0)$, 
etc. This cancellation in the $\theta$ independent and $K_{12}$ 
structures between the 1PI and tadpole contributions originates 
from symmetry relations among vertices summarized in the Ward 
identities. The vanishing of $g_1(k_i,0)$ is a feature of the 
1PI part alone which fits in the requirement of the Goldstone 
theorem to be verified here. All of this also serves as a good 
test of the correctness of the calculation. 

Now we proceed to consider the NC IR behaviour in the total 
self-energies. The $\overline{f_0}$ and $\overline{g_0}$ terms 
are independent of $\theta$, proportional to $p^2$ as in the 
commutative theory and thus IR safe. The $\theta$ dependent 
terms are proportional to some factors quadratic in $p$ due to 
the subtraction, but in principle not necessarily to $p^2$. With 
the constant antisymmetric tensor $\theta_{\mu\nu}$, we may 
construct symmetric and antisymmetric ones by contraction, e.g., 
$\theta_{\mu\nu}^2=\theta_{\mu\rho}\theta_{\nu}^{~\rho}$, which 
may be used to build new NC momenta like 
$\tilde{\tilde{p}}_{\mu}=\theta_{\mu\nu}^2 p^{\nu}$. Thus the 
proportionality factors can be 
$p^2$, $p\cdot \tilde{\tilde{p}}=\tilde{p}^2$, etc. 
The task here is to show that negative powers of 
these scalars never appear in the final results so that the IR 
safety is guaranteed in the self-energies. We shall show that 
the above quadratic $p$ factors will be multiplied by a leading 
factor of order $\ln\theta^2$ or $\ln\tilde{p}^2$. 
(The $p^2\ln p^2$ behaviour is not new as it already appears in 
commutative theory.) Thus the self-energies are IR safe but 
cannot go to the commutative limit smoothly.

Generally speaking, an integral with $K_1$ or $K_{12}$ will 
be IR safe in the external momentum if it is already convergent 
both superficially and in subgraphs without these factors. 
For divergent integrals either superficially or in subgraphs, 
we must be careful. Let us first consider the $K_{12}$ term. We 
shall demonstrate our calculation by some typical terms in 
$\overline{f_{12}(k_i,p)}$. The case of $g_{12}$ is similar but 
much more complicated due to the momentum shifts which introduce 
$p$ in many propagators. Using 
$\overline{A_1(p)A_2(p)}
=A_1(p)\overline{A_2(p)}+\overline{A_1(p)}A_2(0)$, etc, 
we can always subtract sequentially, so that the only complication 
in $g_{12}$ lies in the polynomial $p$ dependence of Feynman 
parameter integrals. Concerning $f_{12}$, the contributions from 
Figs. $(4a)$ and $(4b)$ are cancelled by those of the tadpole. 
Fig. $(4g)$ is finite superficially and in subgraphs without 
$K_{12}$ and is thus safe. The most dangerous is Fig. $(4d)$ 
which has a quadratically divergent subgraph, which in turn may 
transmute into a pole-like singularity in the external momentum. 
Fortunately, this quadratic divergence is cancelled between the 
$\sigma$ and $\pi_0$ contributions proportional to, 
\begin{equation}
\begin{array}{c}
\displaystyle 
\int\frac{d^4k_1}{(2\pi)^4}\int\frac{d^4k_2}{(2\pi)^4} 
D_m^2(k_1)[D_m(k_2)-D(k_2)]\overline{D(k_1+p)}K_{12}.
\end{array}
\end{equation}
When computing loop integrals, we always work on Euclidean 
spacetime to simplify their analytic property. 
(Now $D_m(k)=(k^2+m^2)^{-1}$.) Finishing the $k_2$ integral 
using $I_1$ in appendix B, we have, 
\begin{equation}
\begin{array}{rcl}
{\rm integral}&=&\displaystyle 
-2m^2(4\pi)^{-2}\int_0^1 dx\int\frac{d^4k_1}{(2\pi)^4}
D_m^2(k_1)\overline{D(k_1+p)}K_0
\left(\sqrt{xm^2\tilde{k_1}^2}\right).
\end{array}
\end{equation}
There are preferred directions defined by 
$p_{\mu}$ and $\theta_{\mu\nu}$. To avoid complicated angular 
integration, we have to make some simplifying assumptions 
which should not alter the IR singularity drastically. As 
$\theta^2_{\mu\nu}$ is symmetric and semi-positive definite on 
Euclidean space, it may be diagonalized by an orthogonal 
rotation. We assume that it has a four-fold degenerate 
eigenvalue of $\eta^2$, i.e., 
$\theta^2_{\mu\nu}=\eta^2\delta_{\mu\nu}$ 
with $\eta>0$ a small area scale characterizing NC. Then, 
$\tilde{k_1}^2=\eta^2k_1^2$ and the angular integral is much 
simplified, 
\begin{equation}
\begin{array}{rcl}
\displaystyle 
\int d\Omega_4~(k_1+p)^{-2}&=&\displaystyle 
4\pi\int_0^{\pi}d\omega\sin^2\omega
[k_1^2+p^2+2\sqrt{k_1^2p^2}\cos\omega]^{-1}\\
&=&\displaystyle 
2\pi^2\left\{\begin{array}{rl}
1/k_1^2~(k_1^2\ge p^2)\\
1/p^2~(k_1^2\le p^2)
             \end{array}\right.. 
\end{array}
\label{eqn_ang}
\end{equation}
Including the subtraction $D(k_1)$ term, we are thus 
integrating over $k_1^2\in [0,p^2]$. Note that this seems 
to be a special feature of angular integrals in four dimensions. 
For $p^2\ll m^2$, we have for the dominant $p$ dependent part, 
\begin{equation}
\begin{array}{rcl}
{\rm integral}&\approx&\displaystyle 
m^{-2}(4\pi)^{-4}\int_0^1 dx~
\int_0^{p^2}dk_1^2~k_1^2\left[\frac{1}{p^2}-\frac{1}{k_1^2}\right]
\ln (xm^2\eta^2k_1^2)\\
&\approx&\displaystyle 
-2^{-1}(4\pi)^{-4}m^{-2}p^2\ln(\eta^2m^2p^2), 
\end{array}
\end{equation}
which indeed vanishes as $p\to 0$ but singular as $\theta\to 0$. 
Integrals in Fig. $(4e)$ can be similarly computed to arrive at 
the same conclusion. 

Let us now consider the case when the two loop momenta are 
overlapping. For example, Fig. $(4f)$ contains the 
following integral, 
\begin{equation}
\begin{array}{c}
\displaystyle 
\int\frac{d^4k_1}{(2\pi)^4}\int\frac{d^4k_2}{(2\pi)^4} 
D(k_1+k_2)D_m(k_2)\overline{D_m(k_1+p)}K_{12}.
\end{array}
\label{eqn2K12}
\end{equation}
First finish $k_2$ integral using $I_2$ in appendix B,
\begin{equation}
\begin{array}{rcl}
{\rm integral}&=&\displaystyle 
\int\frac{d^4k_1}{(2\pi)^4}\overline{D_m(k_1+p)}\\
&&\displaystyle \times 
2(4\pi)^{-2}\int_0^1 dx~K_0\left(
\sqrt{(1-x)(k_1^2x+m^2)\tilde{k_1}^2}\right),
\end{array}
\label{eqn_k1}
\end{equation}
where the cosine factor disappears due to $k_1\cdot\tilde{k_1}=0$.
We are interested in the small $p$ limit, 
\begin{equation}
\begin{array}{rcl}
\overline{D_m(k_1+p)}&=&\displaystyle 
D_m(k_1)\left[-2k_1\cdot pD_m(k_1)\right.\\
&&\displaystyle \left.
-p^2D_m(k_1)+4(k_1\cdot p)^2D_m^2(k_1)+O(p^3)\right].
\end{array}
\end{equation}
The first term vanishes since we cannot make up a vector out of 
$\theta_{\mu\nu}$ and $\delta_{\mu\nu}$. It also vanishes by the 
$k_1\to -k_1$ symmetry. The third term is an integral involving 
$k_{1\mu}k_{1\nu}$ which may be simplified as follows. The result 
of the integral is composed of the 
$\delta_{\mu\nu}$, $\theta_{\mu\nu}$, $\theta_{\mu\nu}^2$, 
$\theta_{\mu\nu}^3$, ... terms, where odd products of $\theta$ 
actually cannot appear due to symmetry. As we are also interested 
in the small $\theta$ limit, the $\theta^2$ and higher terms may 
be dropped; namely, we may replace 
$k_{1\mu}k_{1\nu}$ by $k_1^2\delta_{\mu\nu}/4$. 
For $\theta_{\mu\nu}^2=\eta^2\delta_{\mu\nu}$, the above argument 
becomes exact since only one structure $\delta_{\mu\nu}$ is 
possible. We continue to work in this assumption so that the 
angular integration may be finished explicitly. Together with the 
second term, we have $\overline{D_m(k_1+p)}\to -m^2p^2D_m^3(k_1)$. 
To help identify the relevant integration region, we first 
rescale $\eta k_1^2=y$ so that a small $\eta$ will not interfere 
with a large $k_1^2$ in $K_0$. We have, 
\begin{equation}
\begin{array}{rcl}
{\rm integral}&\approx&\displaystyle 
-2(4\pi)^{-4}p^2\epsilon\int_0^{\infty}dy~y(y+\epsilon)^{-3}
\int_0^1dx~K_0\left(\sqrt{(1-x)(xy+\epsilon)y}\right),
\end{array}
\end{equation}
with $\epsilon=m^2\eta$. As $K_0$ decays exponentially at 
a large argument and explodes at a small one, only the latter 
region is important. After some calculation, we obtain the 
following leading term, 
\begin{equation}
\begin{array}{rcl}
{\rm integral}&\approx&\displaystyle 
2^{-1}(4\pi)^{-4}p^2\ln(m^4\eta^2),
\end{array}
\end{equation}
which is IR finite but singular as $\theta\to 0$.
Fig. $(4b)$ in the neutral Goldstone case is convergent without 
the $K_{12}$ factor after it is once subtracted at $p=0$, and 
thus safe in the IR limit. Figs. $(4c,h)$ are similarly done. 
More examples of integrals, especially those appearing in Figs. 
$(4a)$ with many massless propagators that may cause an IR 
problem in the virtual loop momentum, are given in appendix B. 

The neutral Goldstone boson self-energy has an additional 
contribution containing $K_1$. Since $K_1$ involves only 
one of the loop momenta $k_1$, there is a big difference between 
overlapping and non-overlapping integrals. For the latter, it is 
easy to identify the leading singular terms since they are 
essentially a product of two one-loop integrals. Fig. $(4d)$ 
belongs to this category. For example, consider the integral, 
\begin{equation}
\begin{array}{c}
\displaystyle 
\int\frac{d^4k_1}{(2\pi)^4}\int\frac{d^4k_2}{(2\pi)^4}
D_m(k_2)D_m^2(k_1)\overline{D(k_1+p)}K_1.
\end{array}
\end{equation}
It is quadratically divergent in $k_2$ as in commutative 
theory, which is not harmful at all to the Goldstone theorem. It 
is also regular and vanishes in the NC IR limit. For the 
overlapping case, let us begin with an integral appearing in 
Fig. $(4c)$, 
\begin{equation}
\begin{array}{c}
\displaystyle 
\int\frac{d^4k_1}{(2\pi)^4}\int\frac{d^4k_2}{(2\pi)^4}
D_m^2(k_1)D_m(k_2)D_m(k_1+k_2)\overline{D(k_1+p)}K_1.
\end{array}
\end{equation}
Integration over $k_2$ gives a usual logarithmic UV divergence 
plus a $\ln k_1^2$ term. However, the resulting $k_1$ integral 
is again regular and vanishes in the limit of $p\to 0$.  
Fig. $(4e)$ is proportional to the following integral, 
\begin{equation}
\begin{array}{rl}
&\displaystyle 
\int\frac{d^4k_1}{(2\pi)^4}\int\frac{d^4k_2}{(2\pi)^4}
D_m(k_1+k_2)D_m(k_2)\\
&\times
[D(k_1+k_2+p)-D(k_1+k_2-p)]D(k_2+p)K_1\\
=&\displaystyle 
\int\frac{d^4k_1}{(2\pi)^4}\int\frac{d^4k_2}{(2\pi)^4}
D_m(k_1+k_2)D_m(k_2)\\
&\times 
[D(k_2-p)-D(k_2+p)]D(k_1+k_2-p)K_1. 
\end{array}
\end{equation}
It becomes, using $I_2$ in appendix B, 
\begin{equation}
\begin{array}{rcl}
{\rm integral}&=&\displaystyle 
2(4\pi)^{-2}\int_0^1 dx~ 
K_0\left(\sqrt{(1-x)(m^2+p^2x)\tilde{p}^2}\right)\\
&&\displaystyle \times 
\int\frac{d^4k_2}{(2\pi)^4}D_m(k_2)[D(k_2-p)-D(k_2+p)]
\cos(k_2\cdot\tilde{p}), 
\end{array}
\end{equation}
which vanishes by $k_2\to -k_2$.

Now consider overlapping integrals arising in Figs .$(4f,h)$ 
which are nontrivial in the NC IR limit. For example, Fig. $(4f)$ 
contains the following one, 
\begin{equation}
\begin{array}{rl}
&\displaystyle 
\int\frac{d^4k_1}{(2\pi)^4}\int\frac{d^4k_2}{(2\pi)^4}
D(k_1)D_m(k_1+k_2)\overline{D_m(k_2+p)}K_1\\
=&\displaystyle 
2(4\pi)^{-2}\int\frac{d^4k_2}{(2\pi)^4}
\overline{D_m(k_2+p)}\\
&\displaystyle 
\times\int_0^1 dx~
K_0\left(\sqrt{x(m^2+(1-x)k_2^2)\tilde{p}^2}\right)
\cos(xk_2\cdot\tilde{p}).
\end{array}
\end{equation}
Note that only the small $k_2$ region is important for $K_0$ 
where the cosine factor may be ignored for the leading term. 
A similar calculation to eqn. $(\ref{eqn_k1})$ leads to, 
\begin{equation}
\begin{array}{rcl}
{\rm integral}&\approx&\displaystyle 
2^{-1}(4\pi)^{-4}p^2\ln(m^2\tilde{p}^2). 
\end{array}
\end{equation}
And for the purely massless case, we obtain, 
\begin{equation}
\begin{array}{rl}
&\displaystyle 
\int\frac{d^4k_1}{(2\pi)^4}\int\frac{d^4k_2}{(2\pi)^4}
D(k_1)D(k_1+k_2)\overline{D(k_2+p)}K_1\\
\approx&\displaystyle 
2^{-1}(4\pi)^{-4}p^2\ln(p^2\tilde{p}^2).
\end{array}
\end{equation}
Integrals from Fig. $(4h)$ can be similarly computed whose results 
are presented in appendix B. Figs. $(4a,b)$ have no 
contributions containing $K_1$ after subtraction while Fig. 
$(4g)$ is regular in the limit of $p\to 0$. 

\section{Conclusion}

The scalar theory is UV quadratically divergent whether or not its 
symmetry is spontaneously broken. On NC spacetime this virtual UV 
quadratic divergence may transmute into a pole-like IR singularity 
in external momenta. If this occurence persists in scalar theory 
with spontaneous symmetry breaking, it may spoil the validity of 
the Goldstone theorem which is utilized to generate mass through 
the Higgs mechanism when the symmetry is gauged. On naive grounds 
there is no reason why this should not happen. This is especially 
the case when one goes beyond one loop level where the richer 
structure in the NC parameter $\theta_{\mu\nu}$ may produce NC IR 
singularities not appearing at one loop 
$\cite{campbell}\cite{ruiz}$ and the singularities may even be 
enhanced by virtual massless Goldstone bosons in the extra loop. 

We have made a complete analysis of the above problem at two 
loop level by studying the self-energies of the Goldstone bosons 
in the NC $U(N)$ linear $\sigma$ model. We found that the integrands 
in loop integrals have three types of $\theta$ dependence, i.e., 
$\theta$ independent, involving the two loop momenta 
($K_{12}=\cos(2k_1\wedge k_2)$), and involving one loop and one 
external momentum ($K_1=\cos(2k_1\wedge p)$). 
Our crucial observation is that the form factors of the above 
structures vanish in the limit of $p\to 0$. This implies that they 
are effectively once subtracted at $p=0$. The subtraction arising 
from symmetry relations in 1PI and tadpole contributions cancels 
the most singular terms that are harmful to the theorem leaving 
behind a result proportional to a quadratic form in $p$. We have 
computed in detail the leading terms in the coefficients of the 
above form. We observed that delicate cancellation also occurs 
between the Higgs and Goldstone bosons that prevents harmful terms 
in the coefficients. The masslessness of virtual Goldstone bosons 
is not a problem; its IR behaviour can always be separated from 
the one induced by NC. The final leading IR terms in the Goldstone 
self-energies induced by NC are of order $p^2\ln\theta^2$ and 
$p^2\ln\tilde{p}^2$ so that the Goldstone theorem still holds true 
at two loop level. Since the basic mechanism for this mild IR 
behaviour originates from symmetry relations amongst vertices 
of the Higgs and Goldstone bosons, it seems rather natural to 
expect that the theorem should also be valid beyond two loop level. 
On the other hand, the limit of $\theta\to 0$ cannot be smooth at 
two loops and beyond, and this nonsmooth behaviour in $\theta$ is 
not necessarily associated with the IR limit of the external 
momentum as we saw in the leading term $p^2\ln\theta^2$ from the 
$K_{12}$ part.   

{\bf Acknowledgements}

I would like to thank K. Sibold for many helpful discussions 
and for reading the manuscript carefully. 

\appendix

\section{Feynman rules}

For completeness, we list below the Feynman rules for the 
vertices in the noncommutative $U(N)$ linear $\sigma$ model 
with the scalar field in the fundamental representation, which 
were first given in Ref. $\cite{campbell}$. All momenta are 
incoming and shown in the parentheses of the corresponding 
particles. There are no changes in progagators. 
\begin{equation}
\begin{array}{rcl}
\sigma\sigma(p_1)\sigma(p_2)&=&\displaystyle
-i6\lambda v~c_{12}\\
\sigma\pi_0(p_1)\pi_0(p_2)&=&\displaystyle
-i2\lambda v~c_{12}\\
\sigma\pi^+_i(p_1)\pi_j(p_2)&=&\displaystyle
-i2\lambda v\delta_{ij}~e_{12}\\
\sigma(p_1)\sigma(p_2)\sigma(p_3)\sigma(p_4)&=&
-i2\lambda(c_{12}c_{34}+c_{31}c_{24}+c_{23}c_{14})\\
\pi_0(p_1)\pi_0(p_2)\pi_0(p_3)\pi_0(p_4)&=&
-i2\lambda(c_{12}c_{34}+c_{31}c_{24}+c_{23}c_{14})\\
\sigma(p_1)\sigma(p_2)\pi_0(p_3)\pi_0(p_4)&=&
-i2\lambda(2c_{12}c_{34}-c_{13,24})\\
\sigma(p_1)\sigma(p_2)\pi^+_i(p_3)\pi_j(p_4)&=&
-i2\lambda\delta_{ij}~c_{12}e_{34}\\
\pi_0(p_1)\pi_0(p_2)\pi^+_i(p_3)\pi_j(p_4)&=&
-i2\lambda\delta_{ij}~c_{12}e_{34}\\
\sigma(p_1)\pi_0(p_2)\pi^+_i(p_3)\pi_j(p_4)&=&
-i2\lambda\delta_{ij}~s_{12}e_{34}\\
\pi^+_i(p_1)\pi^+_j(p_2)\pi_k(p_3)\pi_l(p_4)&=&
-i2\lambda[\delta_{ik}\delta_{jl}e_{13}e_{24}
          +\delta_{il}\delta_{jk}e_{14}e_{23}]\\
\end{array}
\end{equation}
where the following notations are used:
$p\wedge q=\theta_{\mu\nu}p^{\mu}q^{\nu}/2,~
c_{ij}=\cos(p_i\wedge p_j),~
s_{ij}=\sin(p_i\wedge p_j),~
c_{ij,kl}=\cos(p_i\wedge p_j+p_k\wedge p_l),~
s_{ij,kl}=\sin(p_i\wedge p_j+p_k\wedge p_l)$ and 
$e_{ij}=\exp(-ip_i\wedge p_j)$.

\section{Some examples of two loop integrals involving $\theta$}

We work on Euclidean spacetime where the integrals have a 
simpler analytic property. We start with the one loop integrals 
that have been computed by many authors in the literature. 
\begin{equation}
\begin{array}{rcl}
I_1(\rho^2,m)&=&\displaystyle 
\mu^{4-n}\int\frac{d^nk}{(2\pi)^n}(k^2+m^2)^{-2}\cos(k\cdot \rho),
\end{array}
\end{equation}
where $\rho_{\mu}$ will be identified later with 
$\theta_{\mu\nu}q^{\nu}$ with $q$ a loop or external momentum. 
Using the Schwinger parameter integral to exponentiate the 
denominator, completing the square and shifting $k$, we have
\begin{equation}
\begin{array}{rcl}
I_1&=&\displaystyle 
\int_0^{\infty}d\alpha~\alpha~\mu^{4-n}\int\frac{d^nk}{(2\pi)^n}
\exp[-\alpha(k^2+m^2)+ik\cdot\rho]\\
&=&\displaystyle 
\int_0^{\infty}d\alpha~\alpha
\exp\left[-\alpha m^2-\frac{\rho^2}{4\alpha}\right]~
\mu^{4-n}\int\frac{d^nk}{(2\pi)^n}
\exp[-\alpha k^2]\\
&=&\displaystyle 
\int_0^{\infty}d\alpha~\alpha
\exp\left[-\alpha m^2-\frac{\rho^2}{4\alpha}\right]~
\frac{\mu^{4-n}}{(4\pi\alpha)^{n/2}}\\
&=&\displaystyle 
2(4\pi)^{-2}\left[\frac{4\pi\mu^2}{m^2}\frac{1}{2}
\sqrt{m^2\rho^2}\right]^{2-n/2}K_{n/2-2}(\sqrt{m^2\rho^2}),
\end{array}
\end{equation}
where the remaining parameter integral has been expressed in terms 
of the modified Bessel function $\cite{math}$,
\begin{equation}
\begin{array}{rcl}
K_{\nu}(t)&=&\displaystyle 
\frac{1}{2}\left(\frac{t}{2}\right)^{\nu}\int_0^{\infty}d\alpha~
\alpha^{-1-\nu}\exp\left[-\alpha-\frac{t^2}{4\alpha}\right],~
|\arg t|<\frac{\pi}{2},~{\rm Re~}t^2>0.
\end{array}
\end{equation}
For $n=4$, we have $I_1=2(4\pi)^{-2}K_0(\sqrt{m^2\rho^2})$ which is 
finite except for $\rho^2\to 0$ since $K_0(x)\to -\ln x$ as $x\to 0$.
This is the UV/IR mixing; the virtual UV singularity is regularized 
at the cost of introducing an IR singularity in the external momentum.
Consider the case of $m=0$. The virtual IR singularity in $I_1$ may 
be regularized either by a small mass or working in $n$ dimensions. 
In the latter case, a similar calculation leads to 
\begin{equation}
\begin{array}{rcl}
I_1(\rho^2,0)&=&\displaystyle 
(4\pi)^{-2}(\pi\mu^2\rho^2)^{2-n/2}\Gamma(n/2-2)\\
&=&\displaystyle 
(4\pi)^{-2}\left[\Gamma(n/2-2)-\ln(\pi\mu^2\rho^2)+O(n/2-2)\right],
\end{array}
\end{equation}
where the first term is the virtual IR divergence and the second is 
the would-be virtual UV divergence regularized by the non-vanishing 
external momentum $\rho$. More interesting is the case when $\rho$ 
carries the momentum of a second loop involving massless particles 
so that the virtual IR singularity may be enhanced. Using the above 
result we also obtain,
\begin{equation}
\begin{array}{rcl}
I_2(\rho^2,m)&=&\displaystyle 
\mu^{4-n}\int\frac{d^nk}{(2\pi)^n}[(k+p)^2+m^2]^{-2}
\cos(k\cdot \rho)\\
&=&\displaystyle 
I_1(\rho^2,m)\cos(p\cdot \rho).
\end{array}
\end{equation}

In the following we give the integrals appearing in Fig. $(4a)$. 
We shall assume $\theta^2_{\mu\nu}=\eta^2\delta_{\mu\nu}$ 
throughout for simplicity. 
\begin{equation}
\begin{array}{rcl}
I_3(m,m)&=&\displaystyle 
\int\frac{d^4k_1}{(2\pi)^4}\int\frac{d^4k_2}{(2\pi)^4}
D_m^2(k_1)\overline{D_m(k_2+p)}K_{12}\\
&\approx&\displaystyle 
2^{-1}(4\pi)^{-4}p^2\ln(m^4\eta^2),
\end{array}
\end{equation}
which is computed using $I_1$ and the argument employed to 
simplify the $k_1$ integral in eqn. $(\ref{eqn2K12})$. 
Using $I_1$ and eqn. $(\ref{eqn_ang})$, we have  
\begin{equation}
\begin{array}{rcl}
I_3(m,0)&=&\displaystyle 
\int\frac{d^4k_1}{(2\pi)^4}\int\frac{d^4k_2}{(2\pi)^4}
D_m^2(k_1)\overline{D(k_2+p)}K_{12}\\
&\approx&\displaystyle 
-2^{-1}(4\pi)^{-4}p^2\ln(m^2p^2\eta^2).
\end{array}
\end{equation}
Now consider the integral, 
\begin{equation}
\begin{array}{rcl}
I_3(0,m)&=&\displaystyle 
\int\frac{d^4k_1}{(2\pi)^4}\int\frac{d^4k_2}{(2\pi)^4}
D^2(k_1)\overline{D_m(k_2+p)}K_{12}.
\end{array}
\end{equation}
This integral looks dangerous since there is a virtual IR 
singularity in $k_1$ due to masslessness which may be mixed up 
with that coming from the $k_2$ loop to enhance the final IR 
singularity in $p$. The masslessness may be regularized either by 
a small mass or in dimensional regularization. In the first case, 
the result is obtained from $I_3(m,m)$ by setting the first $m$ 
to be the small mass. It is clear that the two IR singularities 
are separated from each other. In the second case, we proceed as 
follows. Using $I_1(\tilde{k_2}^2,0)$ we have 
\begin{equation}
\begin{array}{rcl}
I_3(0,m)&=&\displaystyle 
\int\frac{d^4k_2}{(2\pi)^4}\overline{D_m(k_2+p)}\\
&&\displaystyle \times 
(4\pi)^{-2}\left[\Gamma(n/2-2)-\ln(\pi\mu^2\tilde{k_2}^2)
+O(n/2-2)\right],
\end{array}
\end{equation}
where the first and second terms are respectively from the IR and 
UV regions of $k_1$. Only the second one is of interest here since 
the first is proportional to $p^2$ and thus does not affect our 
main arguments on the Goldstone theorem. Finishing the $k_2$ 
integral as before, we have the contribution from the small $k_2$, 
\begin{equation}
\begin{array}{rcl}
I_3(0,m)&\approx&\displaystyle 
2^{-1}(4\pi)^{-4}p^2\ln(\mu^2m^2\eta^2), 
\end{array}
\end{equation}
the same as we get from $I_3(m,m)$ using the small mass 
regularization. Withour giving further details, we have, 
\begin{equation}
\begin{array}{rcl}
I_3(0,0)&=&\displaystyle 
\int\frac{d^4k_1}{(2\pi)^4}\int\frac{d^4k_2}{(2\pi)^4}
D^2(k_1)\overline{D(k_2+p)}K_{12}\\
&\approx&\displaystyle 
-2^{-1}(4\pi)^{-4}p^2\ln(\mu^2p^2\eta^2)+\cdots, 
\end{array}
\end{equation}
where the dots are the usual terms in commutative theory that 
vanish in dimensional regularization. Following are the examples 
of integrals appearing in Figs. $(4c,e,h)$,
\begin{equation}
\begin{array}{rcl}
I_4&=&\displaystyle 
\int\frac{d^4k_1}{(2\pi)^4}\int\frac{d^4k_2}{(2\pi)^4}
D_m^2(k_1)D_m(k_2)D_m(k_1+k_2)\overline{D(k_1+p)}K_{12}\\
&\approx&\displaystyle 
-2^{-1}(4\pi)^{-4}m^{-4}p^2\ln(m^2p^2\eta^2),\\ 
I_5&=&\displaystyle 
\int\frac{d^4k_1}{(2\pi)^4}\int\frac{d^4k_2}{(2\pi)^4}
D_m(k_1)D_m(k_2)\overline{D(k_1+p)D(k_2+p)}K_{12}\\
&\approx&\displaystyle 
-(4\pi)^{-4}m^{-2}p^2\ln(m^2p^2\eta^2),\\ 
I_6&=&\displaystyle 
\int\frac{d^4k_1}{(2\pi)^4}\int\frac{d^4k_2}{(2\pi)^4}
D_m(k_1)D_m(k_2)D_m(k_1+k_2)\overline{D(k_1+p)}K_{12}\\
&\approx&\displaystyle 
-2^{-1}(4\pi)^{-4}m^{-2}p^2\ln(p^2\eta). 
\end{array}
\end{equation}

In the following we list integrals involving $K_1$ that arise in 
Fig. $(4h)$ and are nontrivial in the NC IR limit.
\begin{equation}
\begin{array}{rl}
&\displaystyle 
\int\frac{d^4k_1}{(2\pi)^4}\int\frac{d^4k_2}{(2\pi)^4}
D_m(k_1)D_m(k_2)D_m(k_1+k_2)\overline{D(k_2+p)}K_1\\
\approx&\displaystyle 
2^{-1}(4\pi)^{-4}m^{-2}p^2\ln(m^2\tilde{p}^2),\\ 
&\displaystyle 
\int\frac{d^4k_1}{(2\pi)^4}\int\frac{d^4k_2}{(2\pi)^4}
D(k_1)D(k_2)D_m(k_1+k_2)\overline{D_m(k_2+p)}K_1\\
\approx&\displaystyle 
2^{-1}(4\pi)^{-4}m^{-2}p^2\ln(m^2\tilde{p}^2),\\ 
&\displaystyle 
\int\frac{d^4k_1}{(2\pi)^4}\int\frac{d^4k_2}{(2\pi)^4}
D_m(k_1)D(k_2)D(k_1+k_2)\overline{D_m(k_2+p)}K_1\\
\approx&\displaystyle 
2^{-1}(4\pi)^{-4}m^{-2}p^2\ln(m^2\tilde{p}^2),\\ 
&\displaystyle 
\int\frac{d^4k_1}{(2\pi)^4}\int\frac{d^4k_2}{(2\pi)^4}
D(k_1)D_m(k_2)D(k_1+k_2)\overline{D(k_2+p)}K_1\\
\approx&\displaystyle 
2^{-1}(4\pi)^{-4}m^{-2}p^2\ln(p^2\tilde{p}^2). 
\end{array}
\end{equation}


\end{document}